\documentclass[11pt]{article}
\usepackage{moriond,epsfig,color}
\usepackage{axodraw}
\definecolor{Blue}{rgb}{0.3,0.3,0.9}
\def\Journal#1#2#3#4{{#1} {\bf #2}, #3 (#4)}

\def\PRL{\em Phys. Rev. Lett.}
\def\PRD{{\em Phys. Rev.} D}

\def\be{\begin{equation}}
\def\ee{\end{equation}}
\def\bea{\begin{eqnarray}}
\def\eea{\end{eqnarray}}
\def\bfg{\begin{figure}}
\def\efg{\end{figure}}

\begin{document}
\vspace*{4cm}
\title{SINGLE TOP RESULTS FROM CDF}

\author{BERND STELZER for the CDF Collaboration}

\address{University of California, Los Angeles,\\
Dept. of Physics, CA 90095, USA}

\maketitle\abstracts{
The CDF Collaboration has analyzed 955~pb$^{-1}$ of CDF II data collected between 
March 2002 and February 2006 to search for electroweak single top quark production
at the Tevatron. We employ three different analysis techniques to search for
a single top signal: multivariate likelihood functions;
neural networks; the matrix element analysis technique. The
sensitivities to a single top signal at the rate predicted by the Standard
Model are 2.1~$\sigma$, 2.6~$\sigma$ and 2.5~$\sigma$, respectively.
The first two analyses observe a deficit of single top-like events and set
upper limits on the production cross section. 
The matrix element analysis observes a 2.3~$\sigma$ single top excess and 
measures a combined t-channel and s-channel cross section of $2.7^{+1.5}_{-1.3}$~pb.
Using the same dataset, we have searched for non-Standard Model production of
single top quarks through a heavy $W^\prime$ boson resonance.
No evidence for a signal is observed. We exclude at the 95 \% C.L. $W^{\prime}$ boson
production with masses of 760 GeV/$c^2$ (790 GeV/$c^2$) in case
the right handed neutrino is smaller (larger) than the mass of the $W^{\prime}$ boson.}

\section{Introduction}
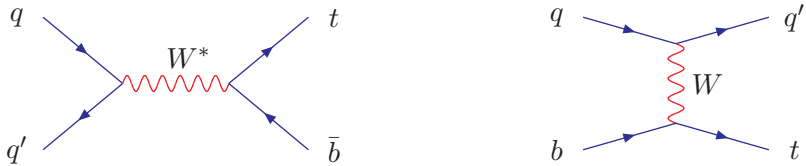
\begin{figure}[h]
\centering
\begin{picture}(140,70)(0,0) 
\Text(-10,50)[]{$q$}
\Text(-10,0)[]{$q'$}
\Text(110,50)[]{$t$}
\Text(110,0)[]{$\bar{b}$}
\SetColor{Blue}
\ArrowLine(0,50)(30,25)
\ArrowLine(30,25)(0,0)
{\SetColor{Red}\Photon(30,25)(70,25){3}{6}}
\Text(55,35)[]{$W^*$}
\ArrowLine(70,25)(100,50)
\ArrowLine(100,0)(70,25)
\end{picture} 
\hspace{2cm}
%do the t-channel
\begin{picture}(80,70)(0,0) 
\Text(-10,50)[]{$q$}
\Text(-10,0)[]{$b$}
\Text(80,50)[]{$q'$}
\Text(80,0)[]{$t$}
\SetColor{Blue}
\ArrowLine(0,50)(35,40)
\ArrowLine(35,40)(70,50)
{\SetColor{Red}\Photon(35,40)(35,10){3}{4}}
\Text(47,25)[]{$W$}
\ArrowLine(0,0)(35,10)
\ArrowLine(35,10)(70,0)
\end{picture} 
\caption{\label{fig:lofeyn}Leading order Feynman diagrams for s-channel (left) and t-channel (right) single top quark production.}
\end{figure}

In 1.96 TeV proton anti-proton collisions at the Tevatron, top quarks are
predominantly produced in pairs via the strong force. In addition, the Standard Model 
predicts single top quarks to be produced through an electroweak t- and s-channel 
exchange of a virtual $W$ boson (Figure \ref{fig:lofeyn}). 
The production cross sections have been calculated at Next-to-Leading-Order (NLO).
For a top quark mass of 175 GeV/$c^2$ the results are 1.98$\pm$0.25~pb and 0.88$\pm$0.11~pb
for the t-channel and s-channel process respectively \cite{harris}. The combined cross section
is about 40\% of the top anti-top pair production cross section. 
The precise measurement of the production cross section allows the direct extraction of
the Cabibbo-Kobayashi-Maskawa matrix element $|V_{tb}|$ and 
offers a source of almost 100\% polarized top quarks \cite{spin}.
Moreover, the search for single top also probes exotic models beyond the
Standard Model. New physics, like flavor-changing neutral currents or heavy 
$W^\prime$ bosons, could alter the observed production rate \cite{anotop}. 
Finally, single top processes result in the same final state as the Standard Model Higgs boson
process $WH\rightarrow Wb\bar{b}$, which is one of the
most promising low mass Higgs search channels at the Tevatron \cite{higgs}.
Essentially, all analysis tools developed for the single top search can be used 
for this Higgs search.

\section{\label{sec:Data}Event Selection}
Our single top event selection exploits the kinematic features of the signal
final state, which contains a real $W$ boson, one or two bottom quarks, and possibly additional
jets. To reduce multi-jet backgrounds, the $W$ originating from the top
quark decay is required to have decayed leptonically. We demand therefore a high-energy
electron or muon ($E_{T}(e)>20$ GeV, or $P_{T}(\mu)>20$ GeV/$c$) and large missing
transverse energy (MET) from the undetected neutrino MET\/$>$25 GeV.
Electrons are measured in the central and in the forward calorimeter, 
$|\eta| <$ 2.0. 
Exactly two jets with $E_{T}>15$ GeV and $|\eta|<2.8$ are required to be present in the event. 
A large fraction of the backgrounds is removed by demanding
at least one of these two jets to be tagged as a $b$-quark jet 
by using displaced vertex information from the silicon vertex detector.
The secondary vertex tagging algorithm identifies tracks associated
with the jet originating from a vertex displaced from the primary
vertex indicative of decay particles from relatively long lived $B$ mesons. 
The backgrounds surviving these selections are $t\bar{t}$, 
$W$ + heavy-flavor jets, i.e. $W+b\bar{b}$, $W+c\bar{c}$, $W+c$ and diboson events $WW$, $WZ$, and $ZZ$.
Instrumental backgrounds originate from mis-tagged $W$ + jets events ($W$ events with light-flavor jets, i.e. with
$u$, $d$, $s$-quark and gluon content, misidentified as heavy-flavor jets) and from non-$W$ + jets events (multi-jet events where one jet is erroneously 
identified as a lepton).

\section{\label{sec:bkg}Background Estimate}
Estimating the background contribution after applying the event selection to
the single top candidate sample is an elaborate process. NLO
cross section calculations exist for diboson and $t\bar{t}$
production, thereby making the estimation of their contribution a relatively
straightforward process. 
The main background contributions are from $W+b\bar{b}$, $W+c\bar{c}$ and $W+c$ + jets, as well as
mis-tagged W + light quark jets. We determine the $W$ + jets normalization from the data and
estimate the fraction of the candidate events with heavy-flavor jets using ALPGEN Monte Carlo samples~\cite{alpgen}, 
which were calibrated against multi-jet data~\cite{jasonprd}. The probability that a W + light-flavor jet
is mis-tagged is parameterized using large statistics generic multi-jet data.
The instrumental background contribution from non-$W$ events is estimated using side-band data with low missing
transverse energy, devoid of any signal, and we subsequently extrapolate the contribution into the signal region with
large missing transverse energy.
The expected signal and background yield in the $W$ + 2 jet sample is shown in Table \ref{njets} and graphically
as a function of $W$ + jet multiplicity next to the table.
\begin{table}
\begin{minipage}{.5\textwidth}
\centering
\begin{tabular}{|lc|}
\hline
Process                 & Number of Events \\
\hline
s-channel	        &  15.4$\pm$2.2	   \\
t-channel               &  22.4$\pm$3.6	   \\
$W + b\bar{b}$		& 170.9$\pm$50.7	     \\
$W + c\bar{c}$		&  63.5$\pm$19.9	     \\
$W + cj$		&  68.6$\pm$19.0	     \\
Mistags			& 136.1$\pm$19.7	     \\
non-$W$	  		&  26.2$\pm$15.9	     \\
Diboson	  		&  13.7$\pm$1.9		     \\
$Z$ + jets              &  11.9$\pm$4.4	 	     \\
$t\bar{t}$              &  58.4$\pm$13.5	     \\
\hline		
Total prediction        & 587.1$\pm$96.6	     \\
\hline		
\hline		
Observed in data	& 644			     \\
\hline
\end{tabular}
\end{minipage}
%\end{table}
%\begin{figure}[h]
\begin{minipage}{.5\textwidth}
\centering \includegraphics[width=.9\textwidth]{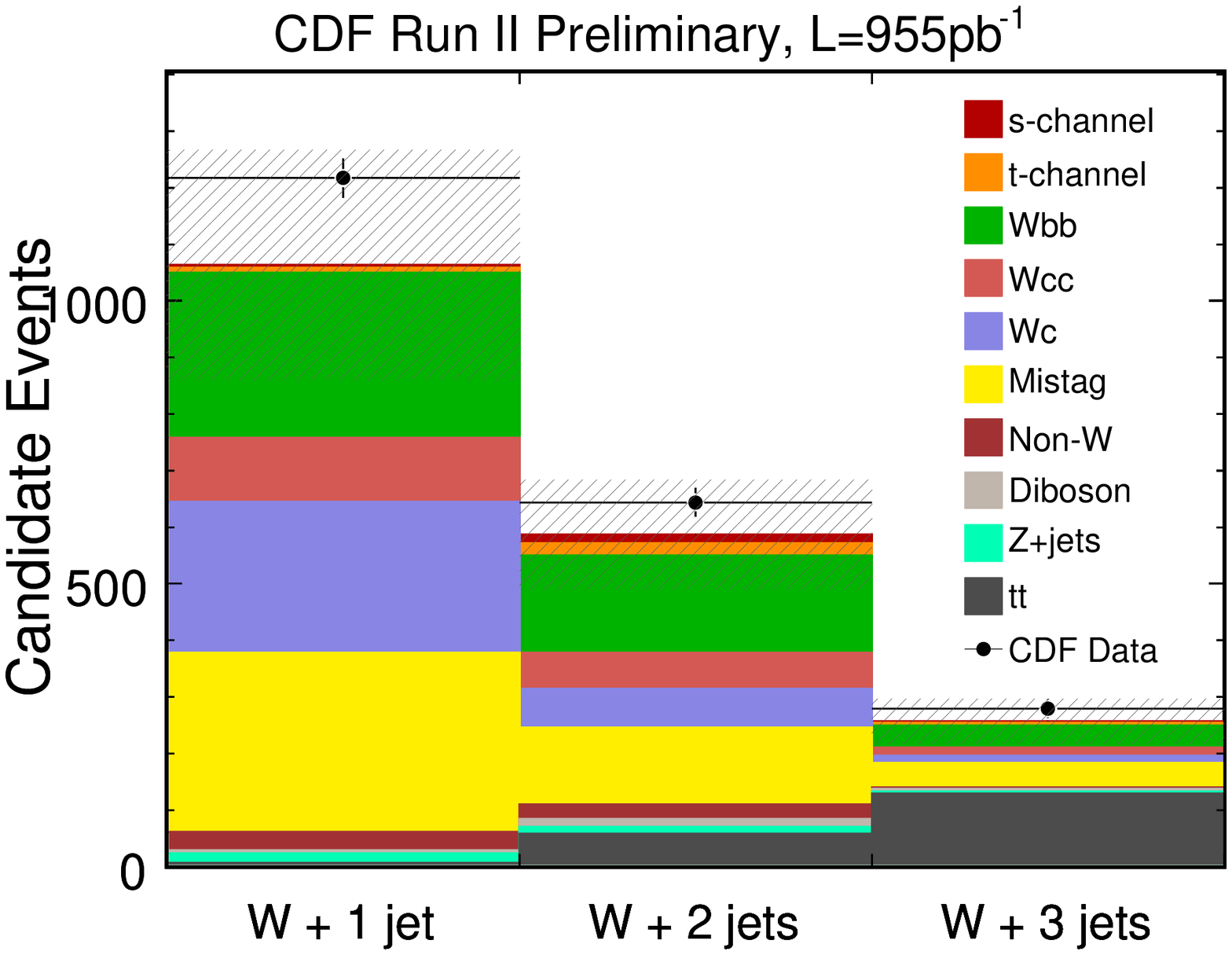}
\end{minipage}
\caption{\label{njets} (Left) Expected signal and background yield in the $W$ + 2 jet sample where at least
on jet is tagged as a $b$-jet. (Right) Graphical sample composition as a function of
the $W$ + jets multiplicity.}
\end{table}
%\end{figure}
Our analysis is performed in the $W$ + 2 jet candidate sample with at least one tagged $b$-jet.
This data sample features the highest signal purity.
Table \ref{njets}, however, demonstrates that the expected amount of single top events
is much less compared to the large amount of expected backgrounds. 
In fact, the uncertainty on the backgrounds is larger than the expected signal, which renders a simple
counting experiment impossible. This is exactly the reason why 
the search for single top quark production requires the best possible
discrimination between signal and background processes and motivates the use
of multivariate analysis tools.

\subsection{Neural Network Jet-Flavor Separation}
Mistags and $W$ + charm events are a large class of background where
no real $b$-quark is present and amount to about 50\% of the $W$ + 2 jets data sample even after 
imposing the requirement that one jet is identified by the secondary vertex $b$-tagger.
We use a neural network tool which uses secondary vertex tracking information to distinguish $b$-jet events
from charm and light-flavor jet events. Figure \ref{fig:nnbtaglf} shows the distribution of this jet-flavor separating 
neural network for the 644 $W$ + 2 jets candidate events. All three single top analyses
use this neural network tool to improve the sensitivity of the analyses.

\section{Analysis Techniques}
No single kinematic distribution encodes all conceivable signal background
separation. We use several sophisticated analysis techniques to combine
information into a single discriminant distribution which is used to extract
the single top content in data.

\subsection{Signal Significance and Discovery Potential}
To quantify the significance of a potential single top signal we use the CLs/CLb method developed at LEP \cite{cls}. 
In this approach, pseudo-experiments are generated from background only events (without single top) 
and from signal plus background events. 
We calculate the probability (p-value) of the background only pseudo-experiments to 
fluctuate to the observed result in data. {\em A-priori} we quote the expected sensitivity to a single 
top signal as the median p-value obtained from the signal + background pseudo-experiments. I.e.
the quantity represents 50\% of the generated signal plus background pseudo-experiments that had a
p-value equal or greater than the expected p-value. All sources of systematic uncertainty are 
included in our statistical treatment and we consider correlation between normalization
and discriminant shape changes due to sources of systematic uncertainty (e.g. the jet-energy-scale uncertainty). 
%The dominant sources
%of systematic uncertainty are the uncertainty on the heayv flavor fraction, the un

\subsection{Multivariate Likelihood Function Analysis}
The multivariate likelihood function analysis computes a joint probability that a
given candidate event originates from signal or background processes given a set of event
characteristics $x_1,..,x_{n_{var}}$. The likelihood ratio, as given in Equation \ref{eq:lf},
is used to build a likelihood function discriminant for s-channel and t-channel single top.
\be
\mathcal{L}(x_1,..,x_{n_{var}}) =\frac{\prod_{i=1}^{n_{var}}p_{sig}^i}{\prod_{i=1}^{n_{var}}p_{sig}^i+\prod_{i=1}^{n_{var}}p_{bkg}^i}
\hspace{1cm}
p_{sig}^i=\frac{N_{sig}^i}{N_{sig}^i+N_{bkg}^i}
\label{eq:lf}
\ee
\bfg[t!]
\centering 
\includegraphics[width=.40\textwidth]{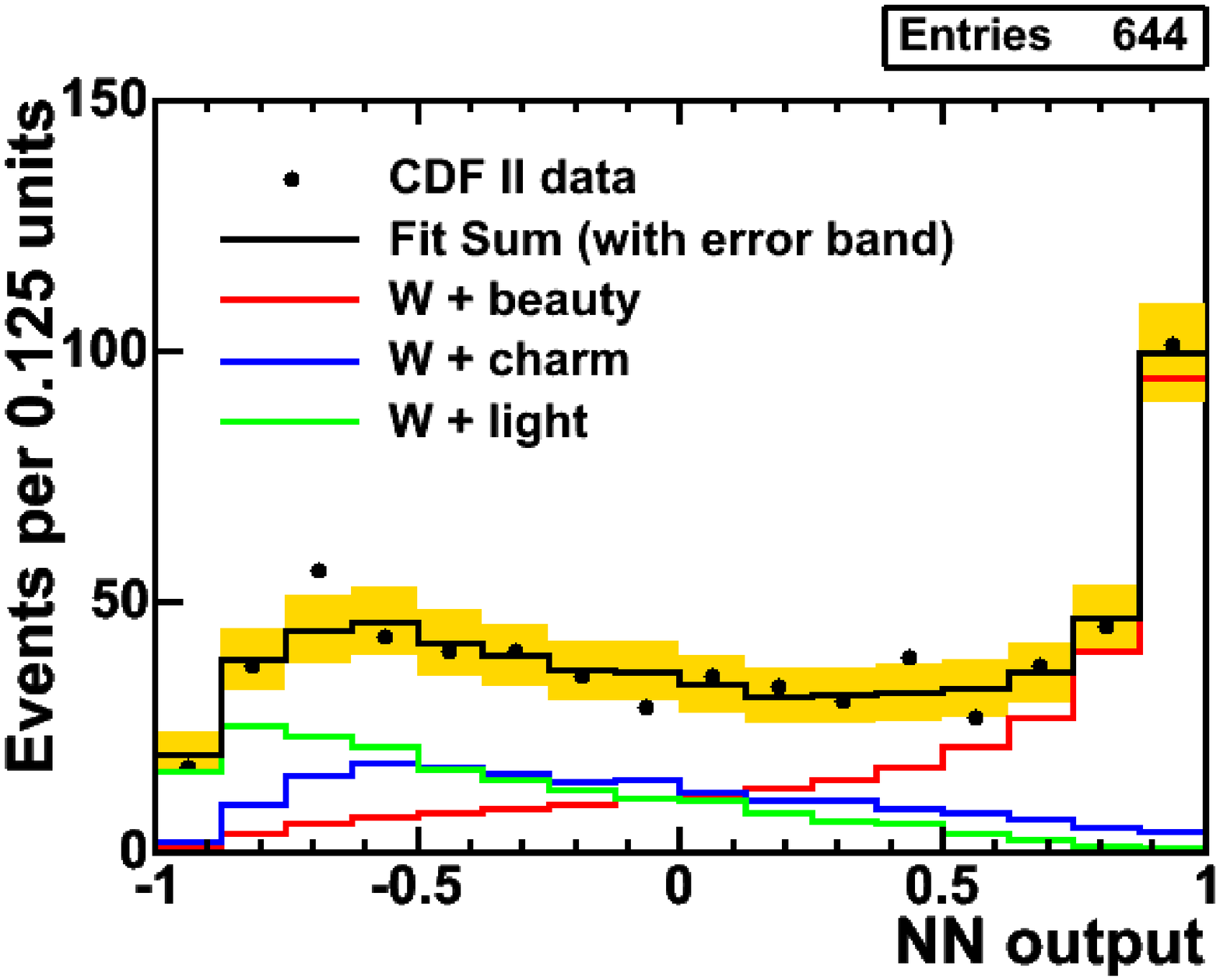}
\hspace{1.5cm}
\includegraphics[width=.34\textwidth]{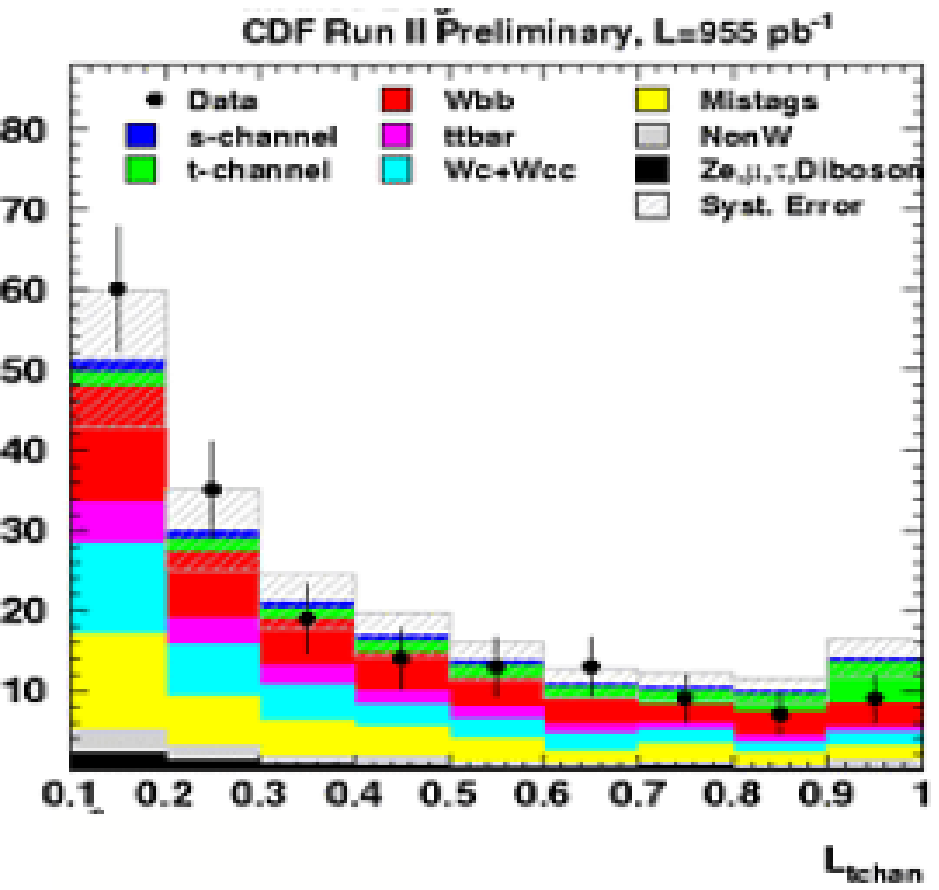}
\caption{\label{fig:nnbtaglf} (Left) Jet-flavor separating neural network applied to $W$ + 2 jets data. (Right) 
Likelihood function discriminant for t-channel single top. The histogram is normalized to the expected
event yield.}
\efg
The t-channel likelihood function discriminant is shown in Figure \ref{fig:nnbtaglf} which used
seven input variables including the jet-flavor separating neural network described in the previous sub-section, 
the mass of the lepton, neutrino and the tagged jet $M_{l\nu b}$, a Matrix Element calculation from 
MadEvent \cite{MadEvent} and other event kinematic variables. The corresponding
s-channel discriminant uses six input variables. The best expected p-value of 2.3\% 
(2.1$\sigma$) is achieved by combining the t-channel and s-channel to a combined single top signal. 
The observed data show no indication of a single top signal and are compatible with a background-only 
hypothesis (p-value 58.5\%). The upper limit on the combined single top cross section is 
2.7 pb at the 95\% C.L. The best fit for the combined s-channel and t-channel
cross sections yields $\sigma_{s+t}=0.3^{+1.2}_{-0.3}$~pb.

\subsection{Neural Network Analysis}
The neural network analysis combines 26 kinematic or event shape related
variables to a discriminant output between -1 for background-like events, and +1
for signal-like events. The five most important input variables are reported by
the neural network package. They are: the output of the jet-flavor separator
neural network, $M_{l\nu b}$, the dijet mass, the pseudo-rapidity
of the untagged jet multiplied by the charge of the detected lepton and
the multiplicity of soft jets in the event with 8 GeV$<E_T<$15 GeV.
To separate t- from s-channel single top quark production, two 
additional networks are trained and a simultaneous fit to both discriminants is performed.
The combined search features an expected p-value of 0.56\% (2.6$\sigma$). The observed
p-value is 54.6\%, providing no evidence for single top production. The corresponding 
upper limit on the cross section is 2.6~pb at the 95\% C.L. The best fit yields
$\sigma_{s+t}=0.0^{+1.2}_{-0.0}$~pb.

\subsection{Matrix Element Analysis}
Using the Matrix Element analysis technique, we compute event-by-event 
probability densities that a given candidate event resulted from a given
underlying interaction (signal or background hypothesis).
The measured four-vectors of the observed jets and the charged lepton serve as experimental input. 
The probability density is computed by integrating over the parton-level
differential cross section $d\sigma$, which includes the leading order matrix element for the
process (calculated using MadEvent \cite{MadEvent}), the parton distribution functions $f(x_i)$, and the detector
resolutions parameterized by transfer functions $W(y,x)$.
Lepton momenta and jet angles are assumed well measured while the jet energy measurements are 
corrected to parton level energies using jet-energy to parton-energy transfer functions.
We integrate over the quark energies and over the $z$-momentum of the neutrino to create a final probability density. 
\be
\label{eq:evtprob}
P(x)=\frac{1}{\sigma}\int d\sigma(y)dq_1dq_2f(x_1)f(x_2)W(y,x)
\ee
We use these probability densities to construct a discriminant
variable for each event (Equation \ref{eq:epd}). We also introduce extra
non-kinematic information by using the output ($b$) of the neural network
jet-flavor separator which assigns a probability ($0 < b < 1$) for each $b$-tagged
jet of originating from a $b$ quark. 
\be
\label{eq:epd}
EPD=\frac{b \cdot P_{single top}}{b \cdot P_{single top}+ b \cdot P_{Wbb} + (1-b) \cdot P_{Wcc} + (1-b) \cdot P_{Wcj}}
\ee
\bfg[t!]
\centering 
\includegraphics[width=.39\textwidth]{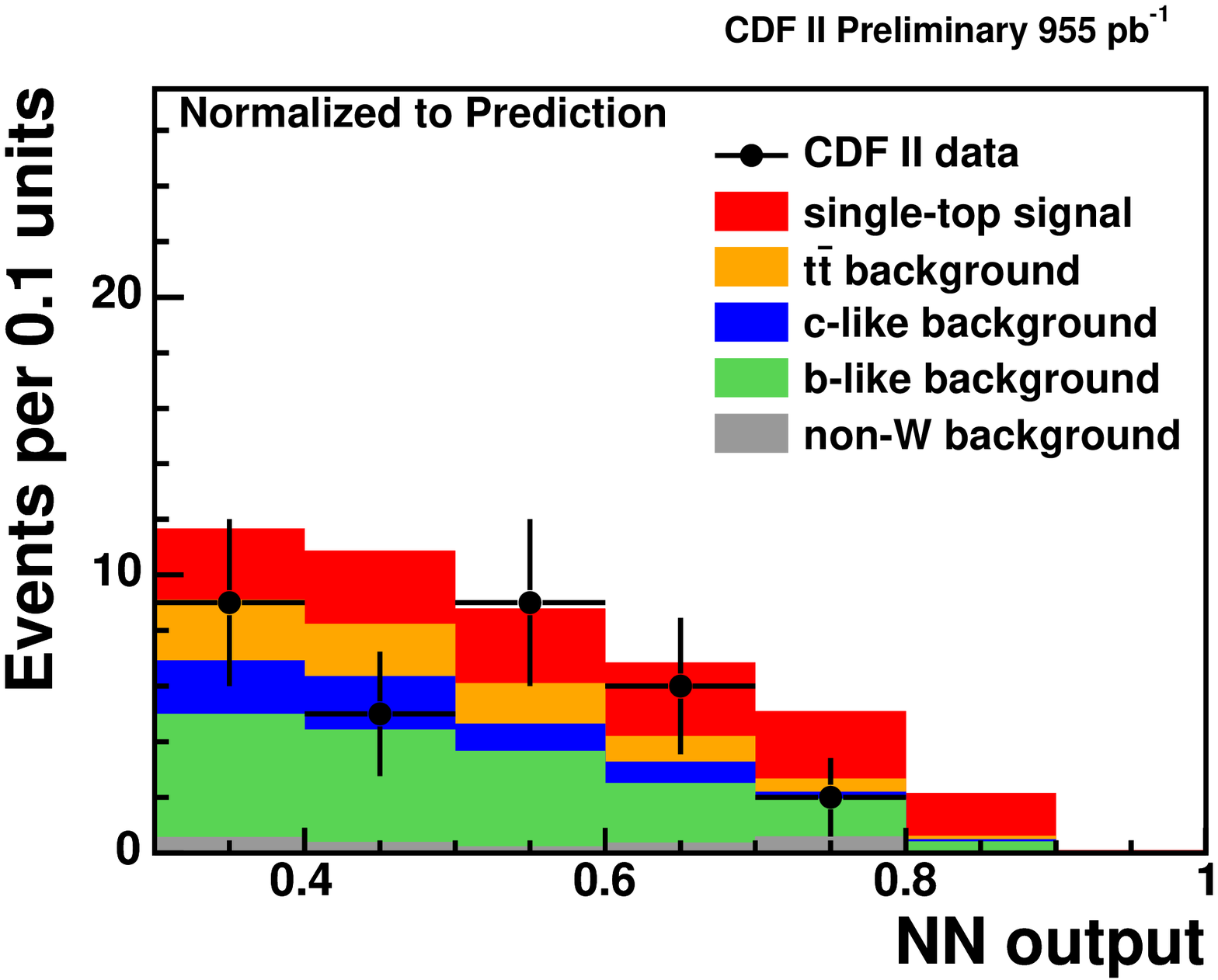}
\hspace{1cm}
\includegraphics[width=.41\textwidth]{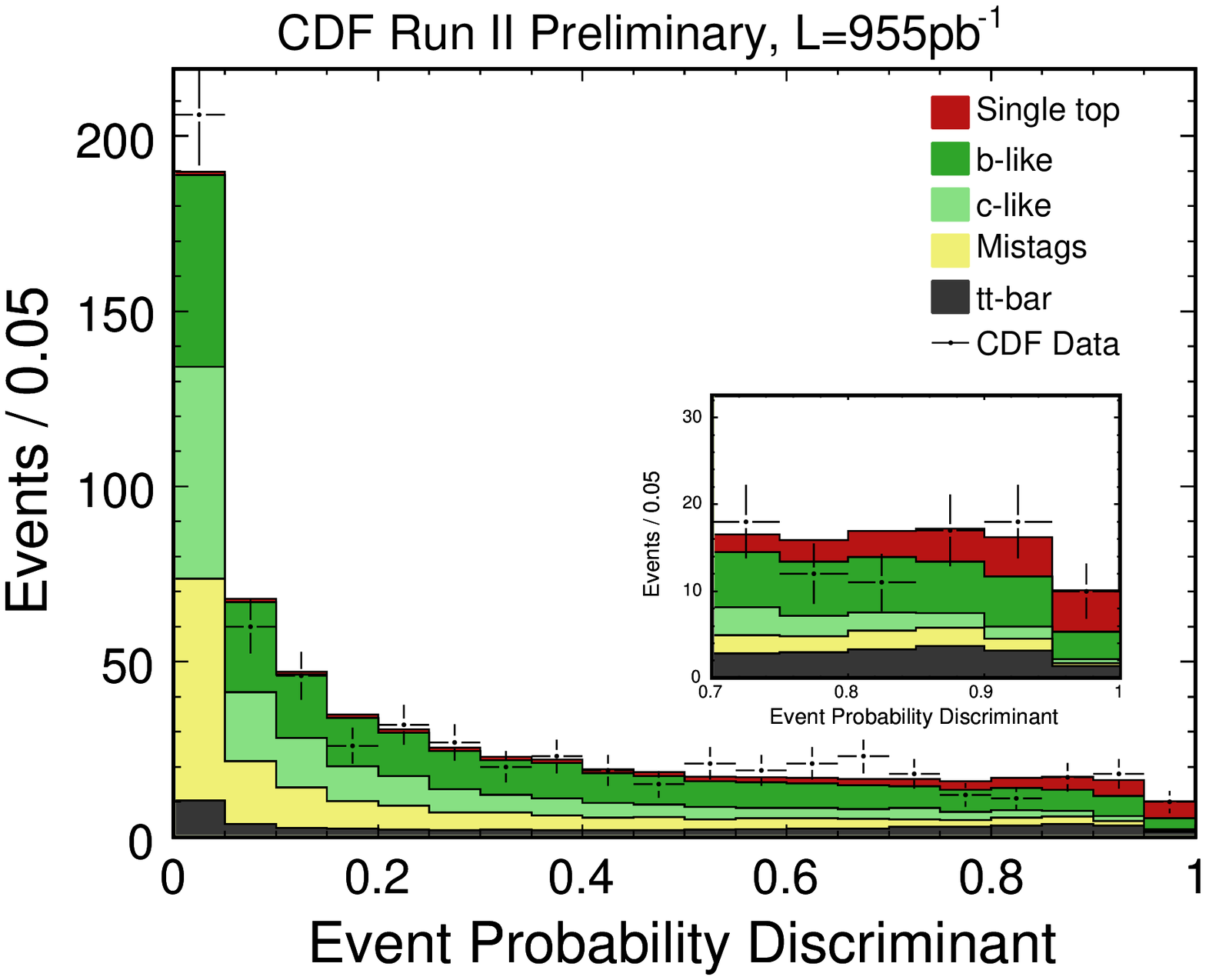}
\caption{\label{fig:epd} (Left) Distribution of the Neural Network analysis discriminant compared to data in the signal 
region (Left). The histogram is normalized to the expected event yield. (Right) Event probability discriminant of the 
Matrix Element analysis. The histogram is normalized to the best fit value. The inset shows the signal region with
high discriminant values.}
\efg
The expected p-value of the combined search is 0.6\% (2.5$\sigma$). The observed
p-value is 1.0\% (2.3$\sigma$), providing a hint for a single top signal. The best fit yields
$\sigma_{s+t}=2.7^{+1.3}_{-1.5}$~pb.

\section{Search for Heavy $W^{\prime}$ Resonances}
Using the same dataset, we have searched for non-Standard Model production of
single top quarks through a heavy $W^\prime$ boson resonance, 
$p\bar{p}\rightarrow W^\prime \rightarrow{t\bar{b}}\rightarrow Wjj$
that appear in models with left-right symmetry, extra dimensions, Little Higgs, and 
topcolor \cite{rhw}.
We look for unexpected structure in the spectrum of the invariant mass of the reconstructed 
$W$ boson and two leading jets ($m_{Wjj}$).
No evidence for resonant $W^\prime$ production is observed and we exclude at the 95 \% C.L. a $W^{\prime}$ 
with Standard Model coupling strength and masses of 760 GeV/$c^2$ (790 GeV/$c^2$) in case
the right handed neutrino is smaller (larger) than the mass of the $W^{\prime}$ boson. 
These new limits exceeds similar searches performed by CDF in Run I and D0 in Run II of the Tevatron 
program \cite{wprimeRun1}.

\bfg[ht!]
\centering
\includegraphics[width=0.40\textwidth]{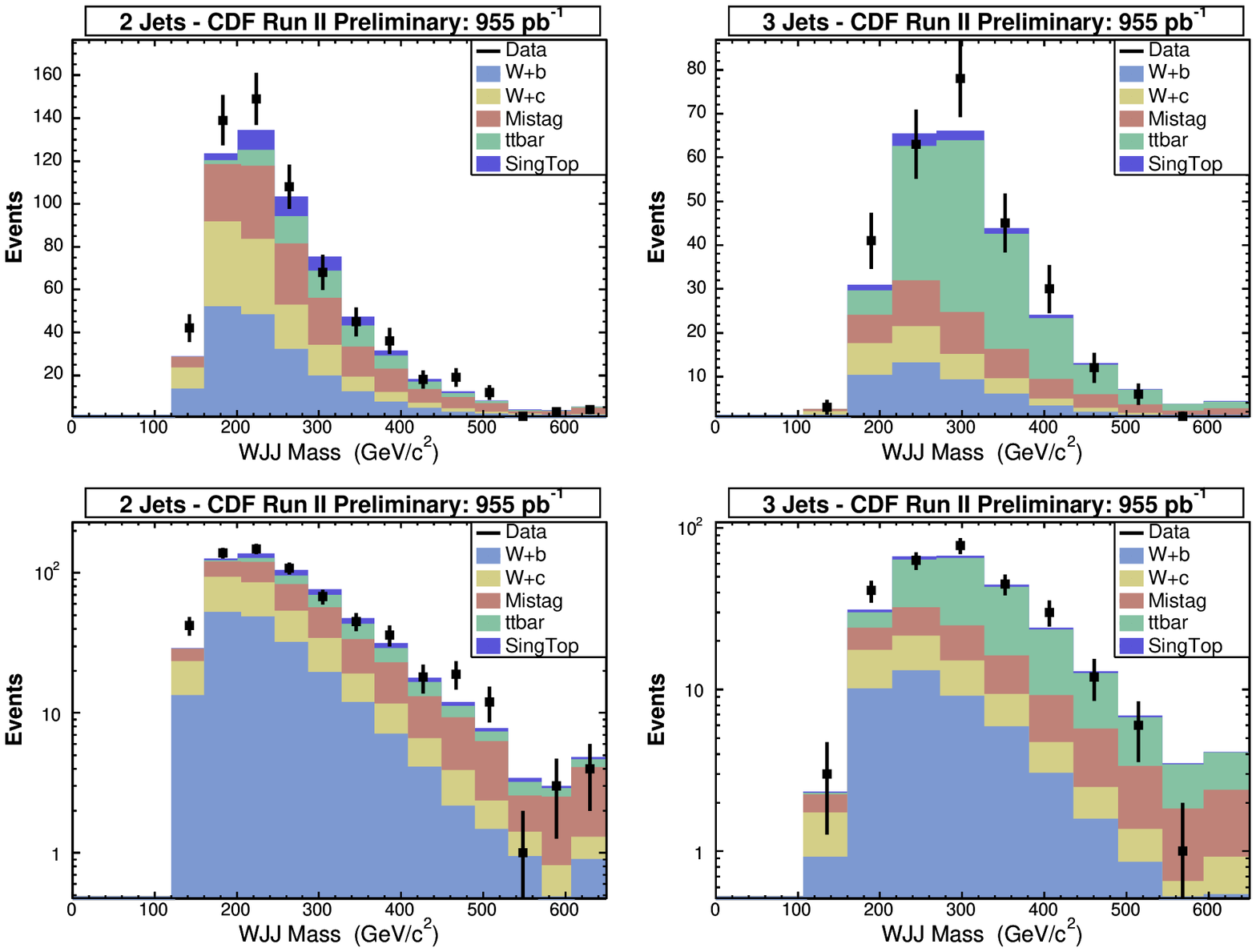}
\hspace{1.5cm}
\includegraphics[width=0.35\textwidth]{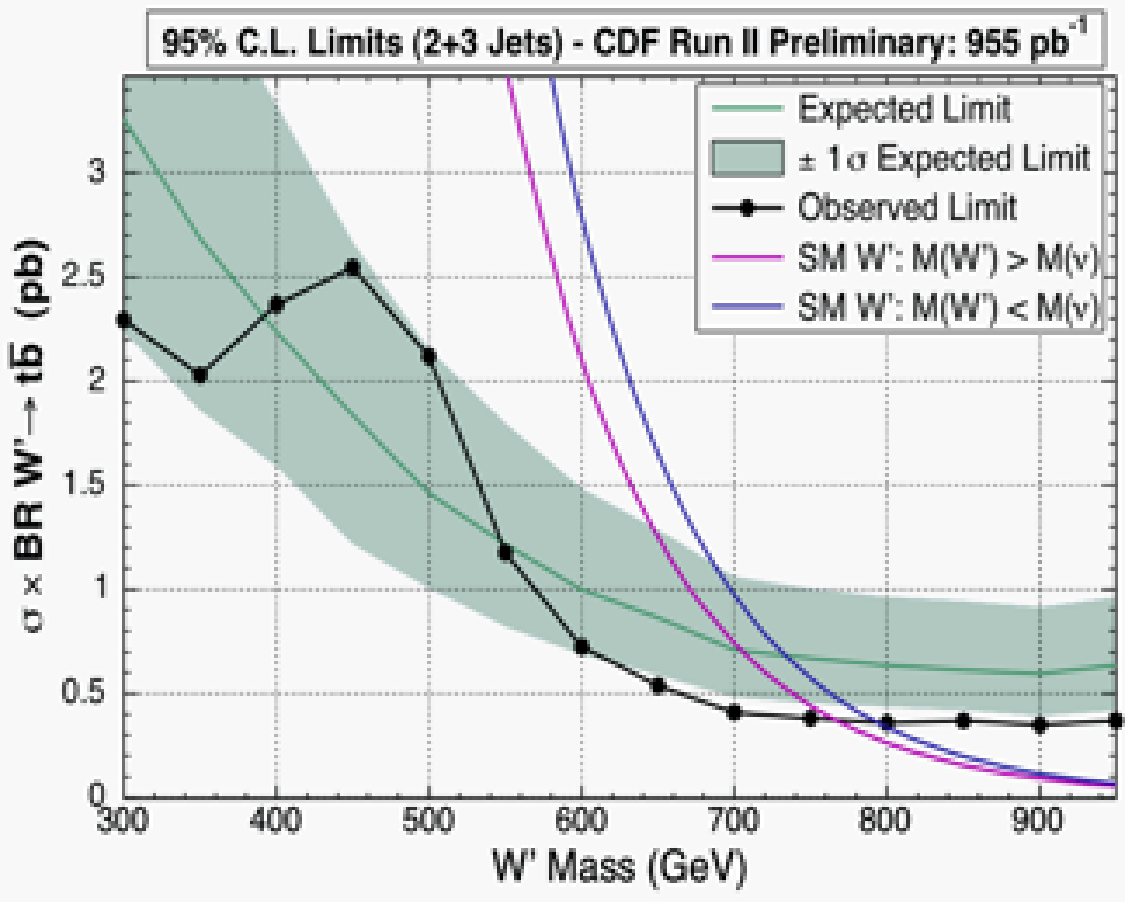} 
\caption{\label{fig:wprime} Background model compared to the data for the $m_{Wjj}$ distribution for $W$ + 2 and 3 jet
events (left). Expected and observed 95\% C.L. limits on the $W^{\prime}$ boson mass assuming Standard Model coupling strength (right).}
\efg
\section{Conclusions}
We have performed searches for electroweak single top quark production
at the Tevatron using 955 pb$^{-1}$ of data collected with the CDF II detector.
The sensitivities to a single top signal at the rate predicted by the Standard
Model range between 2.1 .. 2.6 $\sigma$ and are the most sensitive to-date.
The multivariate likelihood function and neural network analysis observe a
deficit of single top-like events in the data. The matrix element analysis
observes a 2.3 $\sigma$ single top excess consistent with the Standard Model expectation. 
Using pseudo-experiment techniques, we estimated the compatibility of the three analyses to 
about 1.2 \% given the correlation of about 60\% and 70\%. Extensive cross-checks have been
performed to understand the different outcomes in data. At present, there is no evidence for the cause 
other than statistical fluctuations given that the analyses work in different ways and make different, 
analysis specific assumptions.
The larger datasets of 2000 pb$^{-1}$, already
available to the CDF experiment, will clarify what the data are trying to tell us.

Using the same dataset, we exclude at the 95 \% C.L. non-Standard Model single top quark production
through heavy $W^{\prime}$ resonances with masses of 760 GeV/$c^2$ (790 GeV/$c^2$) in case
the right handed neutrino is smaller (larger) than the mass of the $W^{\prime}$ boson.

\section*{Acknowledgments}
I would like to acknowledge the A. v. Humboldt Foundation for supporting this research.

\end{document}